\documentclass[aps,prl,twocolumn,amsmath]{revtex4-1} 

\usepackage{color,graphicx}
\usepackage{dcolumn}
\usepackage{bm}

\usepackage{amsmath}
\usepackage{amsfonts}
\usepackage{amssymb}
\usepackage{rotate}

\usepackage{subfigure}
\usepackage[latin1]{inputenc}

\renewcommand{\-}{\,-\,}

\newcommand{\HC}{\ensuremath{\mathcal{H}_C}}
\newcommand{\HOne}{\ensuremath{\mathcal{H}_1}}
\newcommand{\HPf}{\ensuremath{\mathcal{H}_\text{Pf}}}
\newcommand{\DNF}{\ensuremath{\Delta_\text{NF}}}

\let\oldmarginpar\marginpar
\renewcommand\marginpar[1]{\-\oldmarginpar[\raggedleft\tiny #1]%
{\raggedright\tiny #1}}

\newcommand{\cheatspace}{\vspace*{-10pt}}

\begin{document}

\title{\textbf{Neutral Fermion Excitations in the Moore-Read state at $\nu=5/2$}}

\author{Gunnar M\"{o}ller,$^1$ Arkadiusz W\'ojs,$^{1,2}$ and Nigel R.~Cooper$^1$}
\affiliation{$^1$TCM Group, Cavendish Laboratory, J.~J.~Thomson Avenue, Cambridge CB3 0HE, United Kingdom\\
$^2$Institute of Physics, Wroclaw University of Technology, 50-370 Wroclaw, Poland
}

\date{September 24, 2010}

\begin{abstract}
We present evidence supporting the weakly paired Moore-Read phase in the half-filled second Landau level, focusing on some of the \emph{qualitative} 
features of its excitations. Based on numerical studies, we show that systems with odd particle number at the flux $N_\phi=2N-3$ can
be interpreted as a neutral fermion mode of one unpaired fermion, which is gapped. The mode is found to have two distinct
minima, providing a signature that could be observed by photoluminescence. In the presence of two quasiparticles 
the same neutral fermion excitation is shown to be gapless, confirming expectations for non-Abelian statistics of the Ising model with 
degenerate fusion channels $1$ and $\psi$.
\end{abstract}

\maketitle

Previous studies of the $\nu=5/2$ quantum Hall effect \cite{Willett87,*Eisenstein88,*Pan:1999p411} have accumulated mounting
evidence in favor of the Moore--Read state \cite{MooreRead91,Greiter:1991p1106,*GreiterWenWilczek92,ReadGreen00} of weakly
paired composite fermions (CFs) \cite{Jain89}.
Because of its potential application in topological quantum computation \cite{TQCReview}, scrutinizing
the physical realization of this phase at $\nu=5/2$ is a great challenge of fundamental
and technological importance \cite{Dolev:2008p264,*Radu:2008p419,*Willett:2009p415,TQCReview,Stern:2010p417}.
Theoretically, the main evidence comes from adiabatic connection \cite{MollerSimon08, Morf10,*Wojs:2009p410} and significant
overlaps of the exact ground state in finite model systems with the Moore-Read (MR) state \cite{ReadGreen00} 
or more general weakly paired wave functions \cite{MollerSimon08}.
The key expectation is that that, by extension, the quasiparticles of this gapped
topological phase are described by the same underlying (Ising) conformal field theory
and 
obey non-Abelian statistics.

In this Letter, we provide an approach that tests the \emph{qualitative} properties of the $\nu=5/2$ quasiparticles directly, using
numerical analysis in the spherical geometry \cite{Haldane83}. 
Our results support the predictions of Moore and Read \cite{MooreRead91}, without reference to any trial wave function.
First, as a consequence of the pairing, it is expected that odd-numbered configurations should be disfavored \cite{Greiter:1991p1106}.  
We investigate this effect for a selection of simple two-body Hamiltonians such as the
Coulomb interaction in the second Landau level (LL). (We focus on qualitative features, and exclude 
detailed modeling of effects such as
finite width \cite{PetersonPRL2008} or LL mixing \cite{RezayiSimonMixing2011, BisharaNayak09},
and assume full spin polarization \cite{Morf98,Feiguin09,Wojs10}.) We show
that systems with odd particle number at the flux of the MR ground state possess a dispersing band of low-lying
excitations which we interpret as a neutral fermion (NF) mode arising from an unpaired electron.
This NF mode has an energy gap $\DNF$ of the order of the charge gap $\Delta_c$ \cite{BondersonNeutralFermion}.
Second, we study the energetics of a NF in the presence of charged quasiparticles (QPs):
positive quasiholes (QHs) or negative quasielectrons (QEs).
In this case, our thermodynamic extrapolations of the energy are consistent with a gapless NF. 
This confirms one of the core features of the non-Abelian statistics of the MR
state: the topological degeneracy of two possible fusion channels $1$ and $\psi$ of a pair of two distant QPs, corresponding to the absence 
or presence of an additional fermion.
Furthermore, we determine the NF dispersion and propose an experiment to probe
it directly. We
also give the first evidence that the QHs and QEs of the $\nu=5/2$ state
(for microscopic two-body Hamiltonians) fuse in the $\psi$ channel.

For our studies, we perform exact diagonalizations of model Hamiltonians for $N\leq 20$ spin-polarized, quasi two-dimensional (2D) electrons on a sphere
of radius $R$ pierced by $N_\phi=2N-\sigma$ magnetic flux quanta.
The MR Pfaffian state is at the shift of $\sigma=3$.
We consider three model Hamiltonians: First, the Coulomb interaction $\HC$ in the second LL, as defined by the 
pseudopotential coefficients either for a 2D layer of effective width $w=0$ or $w=3\lambda$ 
($\lambda$ being the magnetic length). 
Second, a modified Coulomb interaction $\HOne$, with the short-range pseudopotential
$V_1$ (for pairs with relative angular momentum $m=1$) increased by $\delta V_1=0.04 e^2/\lambda$.
This increase in $V_1$ is known to yield maximum overlap with the MR Pfaffian \cite{RezayiHaldane00,MollerSimon08}, and was found to mimic 
LL mixing \cite{WojsTokeJainPRL} in the perturbative analysis of Bishara-Nayak \cite{BisharaNayak09,FootNoteLLMixing}.
Third, and finally, the ``Pfaffian (Pf) model'' given by the projector on triplets of minimal relative angular
momentum ($m=3$), $\mathcal{H}_\text{Pf}=\sum_{i,j,k}P_{ijk}^{(3)}$. Though we focus on the Pfaffian state,
the particle-hole symmetry of the two-body interactions $\HC$, $\HOne$
makes our conclusions equally valid for its conjugate, the ``anti-Pfaffian''.

\begin{figure}[bbb]
\begin{center}
\includegraphics[width=0.82\columnwidth]{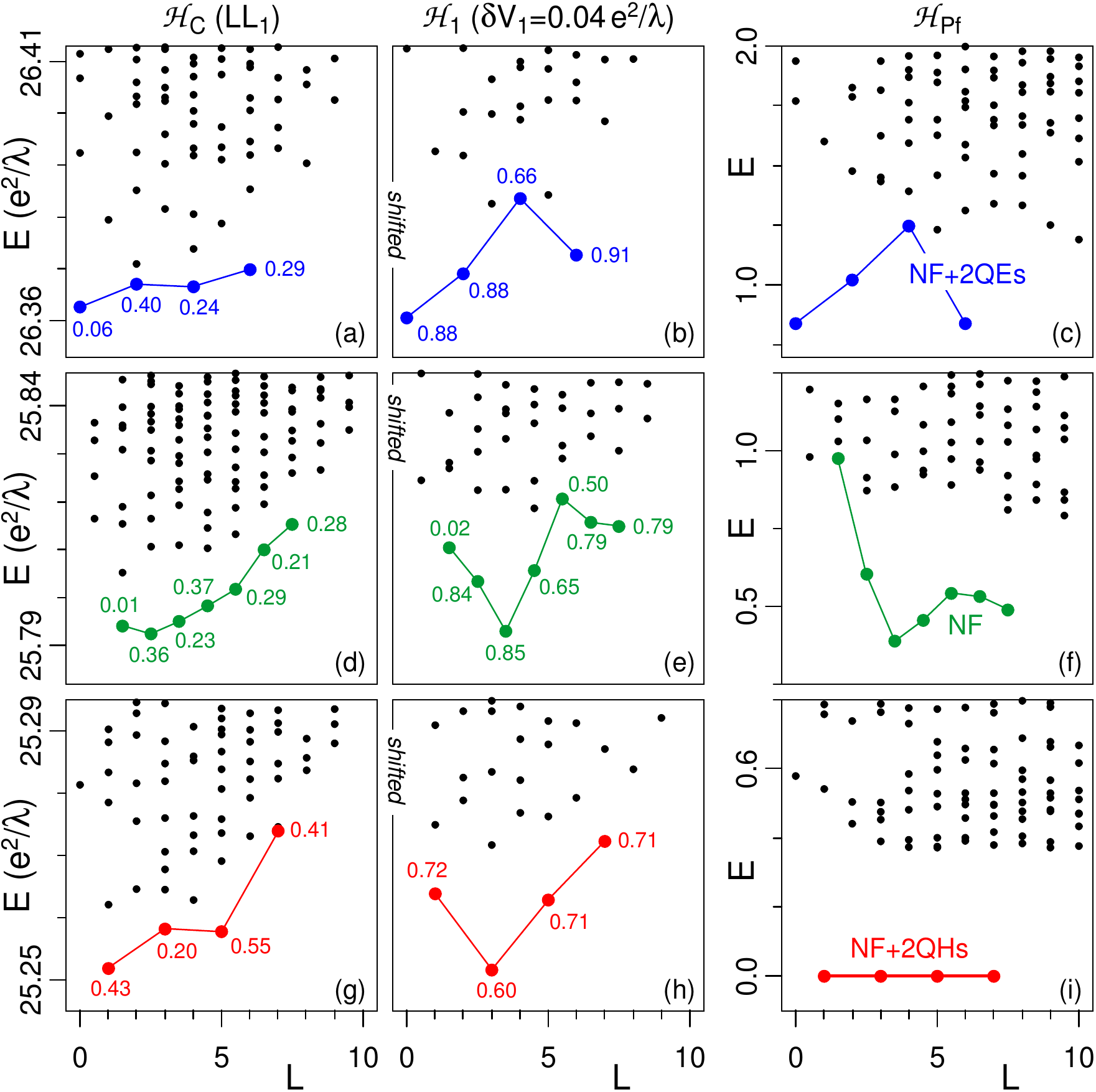}
\caption{(color online) Energy spectra (bare interaction energy $E$ versus angular momentum $L$) of $N=15$ electrons in a half-filled second LL, interpreted in terms of the neutral fermion
(NF) excitation in the Pfaffian (Pf) ground state. Different values of the magnetic flux are: $N_\phi=26$ (top), 27 (center), and 28 (bottom), corresponding to the NF with additional pair of
charged quasielectrons (QEs), NF alone, and NF with additional pair of charged quasiholes (QHs), respectively. Different interactions are: $\HC=$ pure Coulomb (left), $\HOne=$
Coulomb with an additional enhancement of the $m=1$ pair pseudopotential (center), and $\HPf=$ three-body Pfaffian Hamiltonian with the only triplet pseudopotential at $m=3$ (right). 
Labels indicate squared overlaps  $| \langle \mathcal{H}_{C/1}| \HPf \rangle|^2$ for the low-lying bands.
\cheatspace}
\label{fig:Spectra}
\end{center}
\end{figure}

We now analyze the excitation spectra of these Hamiltonians. A typical set of raw data is shown in Fig.~\ref{fig:Spectra} for $N=15$ particles.
Here we first focus on Figs.~\ref{fig:Spectra}(d-f), in which we confirm the existence of a dispersive mode associated with a single fermionic QP 
for finite systems with an odd $N$ \cite{Greiter:1991p1106,*GreiterWenWilczek92}. 
All these three spectra feature a ground state at nonzero angular
momentum within a low-lying band of collective excitations. 
Empirically, we find that this well-separated band extends up to the angular momentum $L=N/2$, as seen most clearly for $\HPf$. 
The minima of the dispersion are far below the continuum.
Eigenstates within the band are spaced by $\Delta L=1$, as expected for a single mobile NF in the background of 
an underlying quantum liquid.

\begin{figure}[bbb]
\begin{center}
\includegraphics[width=0.85\columnwidth]{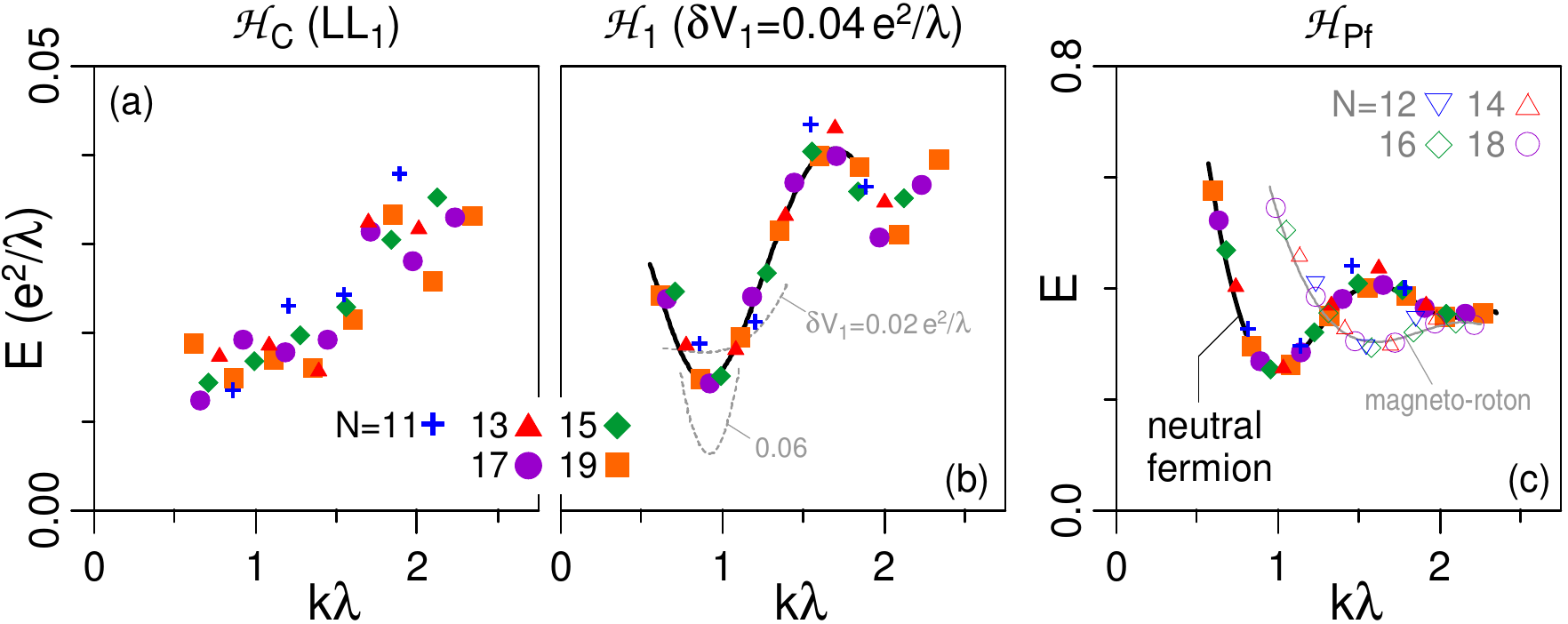}
\caption{(color online) Dispersions (energy $E$ versus wave vector $k$) of the neutral fermion (NF) collective modes of the half-filled second LL, estimated from the systems 
of $N\le19$ electrons at the magnetic flux $N_\phi=2N-3$, for the different Hamiltonians of Fig.~\ref{fig:Spectra}.  Gray dashed lines in (b) show the evolution of \DNF~with $\delta V_1$. 
For comparison, (c) also shows the magnetoroton 
mode for $\HPf$.
\cheatspace}
\label{fig:Dispersion}
\end{center}
\end{figure}

In Fig.~\ref{fig:Dispersion} we collected data from energy spectra for different $N\leq 19$ to estimate the NF dispersion, which is compared
to the magnetoroton 
mode in Fig.~\ref{fig:Dispersion}(c). 
To reduce finite-size effects, for each $N$ we used polynomial interpolation of our 
data to locate the wave vector $k_0$ and the energy $E_0$ corresponding to the minimum of the dispersion.
We find that $k_0(N)$ is essentially constant, and the minimum $E_0$ is well described by $E_0(N) \simeq \DNF + \beta/N$, 
converging to a finite NF gap \DNF~measured with respect to the MR ground state energy \cite{FootNoteGSEnergy}.
Subtracting 
the finite-size scaling, 
our data reduce to one well-defined curve (most accurately for $\HPf$).

The shape of the NF mode varies 
significantly between our model Hamiltonians. A general feature emerging for all spectra is a NF dispersion with two minima [these are best seen in panel (c)]: a 
deeper ``NF$_1$'' near $k_0\lambda \simeq 1$ and a more shallow ``NF$_2$'' near $k\lambda\simeq 2$. The dispersion of 
\HC~shows strong finite-size effects, which we interpret as a consequence of the proximity to a phase transition into a charge-density wave phase
\cite{RezayiHaldane00}. Indeed, \HOne~which is known to yield a state well inside the weakly paired phase also produces cleanly defined dispersions, particularly for the NF. 
(The magnetoroton dispersion for $\HC$ or $\HOne$ -- not shown -- has stronger
finite-size effects than the NF, as it involves two interacting QPs instead of one.)
Comparing different panels, it is remarkable that as soon as the two minima of the NF dispersion actually form (which for $\HOne$ seems to require 
$\delta V_1 \gtrsim 0.02\,\!e^2/\lambda$), they remain located at virtually unchanged wave vectors, while the bandwidth depends significantly on the particular 
model (e.g., on $\delta V_1$).
The 
NF$_1$ lies slightly below the Fermi surface of CFs, {\it i.e.} $k_0 \simeq k_F$ (for a half-filled LL of 
spinless fermions, $k_F = 1/\lambda$). This confirms the expectation of Bogoliubov theory, that in a weakly paired
phase of CFs, and for weak coupling, the minimum is close to $k_F$ \cite{ReadGreen00}. 
The 
presence 
of the second minimum NF$_2$ is more surprising. 
It could arise as a superposition of a NF with additional magnetoroton excitations. 
However, the combined energy for a NF and magnetoroton is found to be larger than NF$_2$. 
Tentatively, this feature could be related to a bound state of these objects.
In any case, we conclude that NF$_2$ cannot decay into a NF and a magnetoroton, so it is a genuine feature of the NF dispersion describing 
a long-lived excitation, and can be tested in experiment.

Direct observation of the NF requires a probe changing the fermion number of the second LL. One such probe is photoluminescence (PL), in which an electron in this LL 
recombines with a photoexcited valence band hole (the `1,0' or `1,1' PL lines in Ref.~\cite{Stern:2010p801}). The PL spectrum depends on the nature of the state into which the 
hole relaxes prior to recombination. Often for fractional quantum Hall systems, this is an ``excitonic'' state in which the hole binds a charge $e$ to form a neutral exciton moving in the 
background incompressible liquid \cite{Cooper:1997p406}. Interestingly, for the $\nu=5/2$ state there are two distinct excitonic states \cite{FootNoteSpinPolarization}
with different parity of the electron number $N_e$ \cite{FootNoteParity}.
Which of these two states has the
lower energy is difficult to predict: this amounts to determining the fusion channel of four QEs in the presence of the hole. 
However, if such fusion can be viewed as pairwise, then the two pairs will fuse either both to $1$ or both to $\psi$, giving an overall even $N_e$. PL
recombination removes one electron, so it leaves a final state with opposite parity to the initial state. For an odd initial $N_e$, recombination can occur to the MR 
ground state, 
yielding a sharp PL line (symmetrically broadened by disorder). For even initial $N_e$, recombination leaves an odd final $N_e$, and so 
\emph{must involve the creation of a NF}.  Since the excitonic state is typically easily localized by disorder \cite{Cooper:1997p406}, the resulting PL spectrum will 
probe the density of states of the NF band. The minima in the NF band will appear as two asymmetrically broadened peaks 
\footnote{For each peak, there can be additional `shake-up' of magnetorotons, leading to threshold peaks at lower energy.}. 
Observation of this double-peak structure in PL would allow direct measurements of the minima of the NF band.

The evolution of the dispersion minimum NF$_1$ as a function of $\delta V_1$, sketched as gray lines in Fig.~\ref{fig:Dispersion}(b), 
gives some insight into the nature of phase transitions. 
At small $\delta V_1$, the NF gap \DNF~remains nonzero, so we expect the pairing nature of the phase to survive up to the transition into a 
charge-density wave \cite{RezayiHaldane00}. 
The collective NF mode flattens at small $\delta V_1$; however, the spectrum is dominated by finite-size effects below $V_1 \approx 0.02$.
At large $\delta V_1$ (approaching interactions resembling the lowest LL) a smooth decay of \DNF~indicates a continuous weakening of pairing (see also \cite{MollerSimon08}). 
The minimum becomes steeper, and the gap collapses near $k\approx k_F=1/\lambda$, consistent with a crossover into the CF Fermi liquid state.

We now turn to investigate the physics of a neutral fermion in the presence of two QEs or QHs. For a pair of QHs in the MR state (even $N$), 
$\HPf$ features a band
of zero-energy states spaced by $\Delta L=2$ and terminating at $L=N/2$ \cite{ReadRezayi96}. We find that when a neutral fermion is added to the system, 
low-lying states are found at the same angular momenta [see Figs.~\ref{fig:Spectra}(g-i)]. Empirically, we find that 
quasielectron states behave similarly. First, we identify a band of low-lying states for $\HPf$ at even $N$, again with $\Delta L=2$ but terminating at $L=N/2-2$. For 2QE+NF 
configurations, a band with the same angular momenta is obtained by \emph{removing} one particle and two flux from the system, so the lowest energy NF 
states can be thought of as holelike.
Figures \ref{fig:Spectra} (a-c) show example 2QE+NF spectra.

\begin{figure}[tbhp]
\begin{center}
\includegraphics[width=0.85\columnwidth]{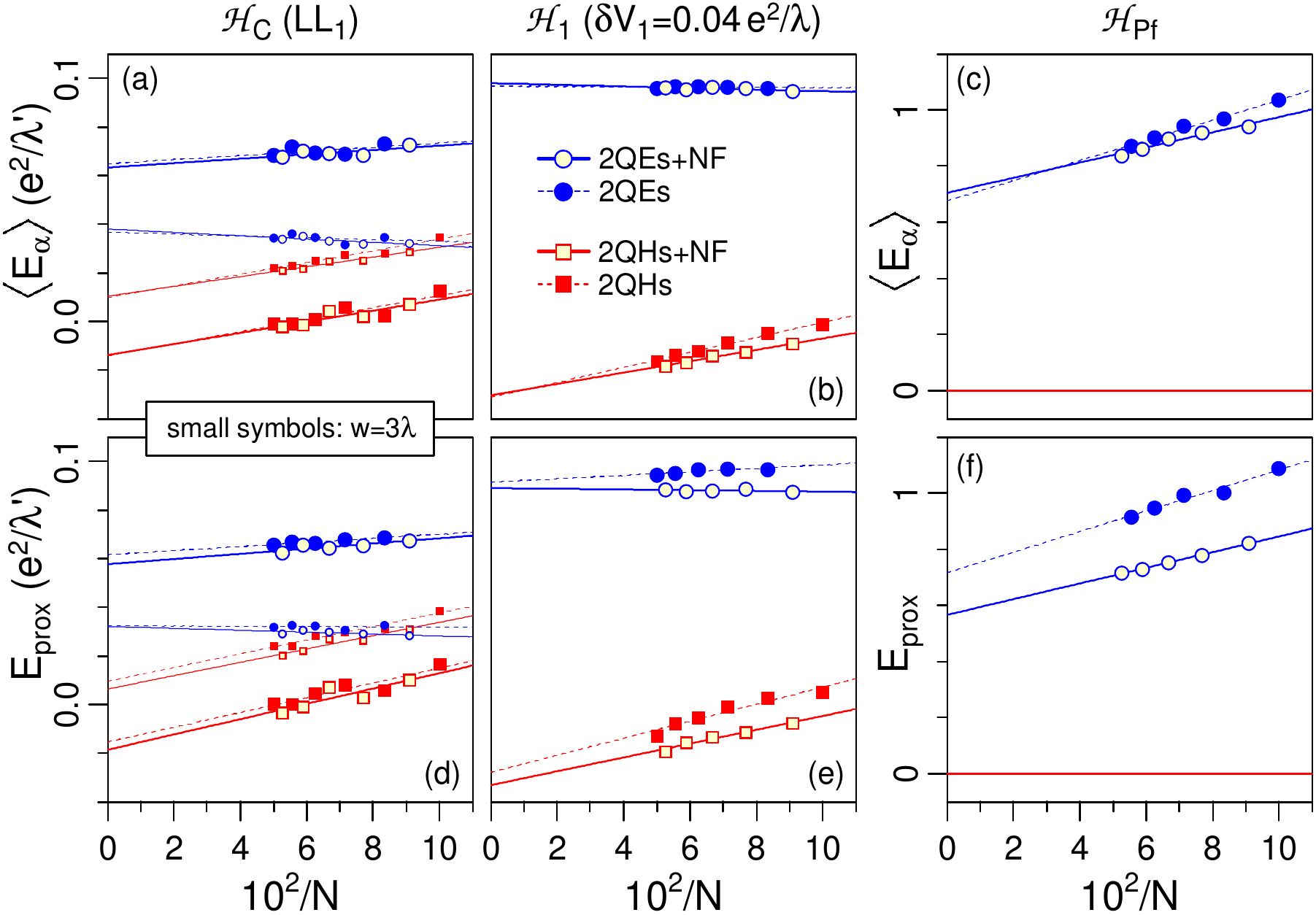}
\caption{(color online) Comparison of the total energies of a pair of QEs or QHs, with and without an additional NF (see text for the precise definition), for systems of different size $N$ and for the different Hamiltonians of Fig.~\ref{fig:Spectra}. Top: energies $\langle E_\alpha \rangle$ averaged over all 2QE/2QH ($+$NF) states; bottom: energies $E_\text{prox}$ taken for the smallest average QE--QE or QH--QH distance. While $\langle E_\alpha \rangle$ extrapolate to similar values with or without NF, the differences are significant for $E_\text{prox}$.
\cheatspace}
\label{fig:Scaling}
\end{center}
\end{figure}

In the presence of QPs the low-lying excitations are not as well separated in the spectrum as for the NF alone, 
and more significant finite-size effects are expected. Thus, we proceed carefully in analyzing the energetics. 
The presence of the NF can affect the angular momentum, relative positions, and shape of the QPs, changing their interaction energy in an unknown way. Since we are unable to 
subtract these effects systematically, instead we average the energy over all states in the low-energy band associated with the two QP(+NF) excitations. 
Thus, we evaluate the (properly normalized) average $\langle E_\alpha \rangle \sim \sum_L (2L+1) E_\alpha(L)$. 
The energy of each eigenstate $E_\alpha(L)$ is measured with respect to the ground state energy \cite{FootNoteGSEnergy} 
(where $\alpha$ indicates a set of QP numbers, $\alpha$=2QE, 
2QH, 2QE+NF, and 2QH+NF). For Coulomb Hamiltonians, we apply standard corrections to the energies $E_\alpha(L)$, including using rescaled magnetic 
length,
applying an electrostatic charging correction of the energies, and correcting for the Coulomb interactions of QPs, $\delta V_\text{QP}$, in the excited 
configurations \cite{Morf02,FootNoteVqp,[{Neither $\langle E_\alpha \rangle$ nor $E_\text{prox}$ represent the definition of the proper quasiparticle energy [}][{.], and thus can be negative.}]MorfHalperinPRL86}.$\phantom{\text{\cite{MorfHalperinPRL86}}}$

The values $\langle E_\alpha \rangle$ obtained after applying the charging corrections are shown in Figs.~\ref{fig:Scaling}(a-c) 
and reflect the total energy of a system with two QEs or QHs, with or without an additional NF \cite{MorfHalperinPRL86}. 
Importantly, for odd $N$ it contains the energy cost for adding a NF in the presence of a pair of QEs ($\Delta_{\rm NF}^-$) 
or QHs ($\Delta_{\rm NF}^+$), that includes the interaction of the NF with these QPs.
Since we average over all positions of the QPs, in finite-size systems one expects a nonzero splitting $\Delta_{\rm NF}^\pm$ between the $\psi$ and $1$ channels from 
configurations with overlapping QPs. Estimates from trial states of QHs at close separation \cite{Baraban:2009p302} 
suggest that for Coulomb interactions the 
splitting is  $\simeq 0.01\,\!e^2/\lambda\equiv \Delta_\text{max}$. 
When averaged over all possible QP positions, the contribution would be significantly smaller,
due in part to its oscillatory behaviour.
For finite systems, we find the (average) splitting of fusion channels, including its finite-size effects, satisfies $\Delta_{\rm NF}^\pm\lesssim\Delta_\text{max}$.
(For $\HPf$ in Fig.~\ref{fig:Scaling}(c), the splitting $\DNF^+$ vanishes by construction \cite{ReadRezayi96}.)

If the QPs are non-Abelian Ising anyons, as in the Moore-Read phase, then the energy splitting $\Delta_{\rm NF}^\pm$ should scale to zero in the thermodynamic limit. 
The extrapolations in Fig. ~\ref{fig:Scaling}(a-c)  are consistent with the vanishing of both splittings $\Delta_{\rm NF}^\pm$ (in each case, the extrapolated value is considerably smaller than its standard deviation).
Although we cannot prove that the splittings vanish exactly, we emphasize that their best estimates are at least an order of magnitude smaller than the charge gap $\Delta_c$ or $\DNF$.
It is highly nontrivial to find a near degeneracy on this scale. 
We have examined the behaviour as a function of $\delta V_1$ in our model Hamiltonian \HOne.
We find that the splitting remains similarly small over the same range of interactions for which the $L=0$ ground state has a large gap and a high overlap with the Moore-Read Pfaffian [11, 12], {\it i.e.}, extending upwards in $\delta V_1$ from about $\HC$ towards the point of collapse of the NF gap. 
We take these results as evidence that, over this range of $\delta V_1$, the QEs and the QHs in these realistic systems have non-Abelian exchange statistics of the form of the Moore-Read phase.

To investigate which fusion channel is preferred at short distance, we have estimated the splitting $\Delta_{\rm NF}^\pm$ for QPs at near-coincident points. 
In Fig.~\ref{fig:Scaling}(d-f), we show $E_\text{prox}=E_\alpha(L_\text{max})$ using the low-lying 2QP / 2QP+NF states with the largest angular 
momentum (and closest separation of the QPs) \footnote{Here, the correction $\delta V_\text{QP}$ was taken as $(q/e)^2\,V_1$.}. 
In this case the odd--even splitting opens for each of the considered Hamiltonians. The splitting also remains in the
thermodynamic limit, revealing a slightly negative $\Delta_{\rm NF}^\pm$, signaling a preference for the $\psi$-channel [e.g., 
$\Delta_{\rm NF}^+=-0.0053(41)\,e^2/\lambda$ and
$\Delta_{\rm NF}^-=-0.0023(23)\,e^2/\lambda$ for $\HOne$, 
and $\Delta_{\rm NF}^+=-0.15(3)$ for $\HPf$]. 
Previously, the splitting had been known only for QHs, and was based on variational wave functions \cite{Baraban:2009p302}.
Here, we report the splittings for both QE and QH based on exact calculations in finite systems.

In conclusion, we have analyzed the neutral fermion excitations of the
$\nu=5/2$ state for microscopic two-body Hamiltonians. We showed that
these exhibit similar properties to those of the Pfaffian model, for
which the Moore-Read phase is the exact ground state. The neutral
fermion is gapped in the ground state, but is gapless in the presence
of a pair of quasiparticles.  These results provide important evidence
that the $\nu=5/2$ state has the properties of the weakly paired
Moore-Read phase. Our studies also elucidate additional physical
properties of this phase: we predict that characteristic features of
the NF dispersion can appear in photoluminescence experiments; and we
show evidence for the nature of the fusion of two QEs or QHs.

\begin{acknowledgments}
We acknowledge support from Trinity Hall Cambridge (G. M.); 
the EU under Marie Curie Grant PIEF-GA-2008-221701 and
MNISW under N202179538 (A. W.); and from
EPSRC under EP/F032773/1 (N. R. C.).
\end{acknowledgments}

\bibliography{five-halfs}

\begin{thebibliography}{39}
\bibitem [{Foo(){\natexlab{b}}}]{FootNoteGSEnergy}%
  \BibitemOpen
  \bibinfo {note} {For odd $N$, we define an equivalent total groundstate
  energy $E_0^\text{eq}(N)$ that is interpolated linearly from the neighbouring
  (even) values $N\pm 1$; $\beta$ depends on the Hamiltonian, e.g.
  $\beta=0.08\,e^2/\lambda$ for \HOne~and $\beta=2$ for \HPf; we define wave
  vectors as $k=L/\sqrt{2(N_\phi-2)}$.}\BibitemShut {Stop}%
\bibitem [{Foo(){\natexlab{c}}}]{FootNoteSpinPolarization}%
  \BibitemOpen
  \bibinfo {note} {We assume initial states are spin-polarized. At weak field,
  excitonic states could involve spin-textures such as skyrmions
  \cite{Wojs10}.}\BibitemShut {Stop}%
\bibitem [{Foo(){\natexlab{d}}}]{FootNoteParity}%
  \BibitemOpen
  \bibinfo {note} {The parity can be defined by counting the number of
  electrons in a region enclosing a localized excitonic wave
  packet.}\BibitemShut {Stop}%
\bibitem [{Foo(){\natexlab{e}}}]{FootNoteVqp}%
  \BibitemOpen
  \bibinfo {note} {For angular momentum eigenstates $\delta V_\text{QP}$ is
  expressed via $V(L)$, the electronic pseudopotentials of the lowest LL
  rescaled by $(q/e)^2$, such that $\langle \delta V_\text{QP}\rangle \propto
  \sum (2L+1) V(L)$.}\BibitemShut {Stop}%
\end{thebibliography}

\end{document}